\documentclass[%
reprint,
superscriptaddress,
 amsmath,amssymb,
 aps,
 prl,
]{revtex4-2}

\usepackage{booktabs}
\usepackage{graphicx}
\usepackage{dcolumn}
\usepackage{bm}
\usepackage[dvipsnames,table]{xcolor}
\usepackage[colorlinks=true,linkcolor=Blue,citecolor=Blue,urlcolor=Blue]{hyperref}
\usepackage[capitalise]{cleveref} 

\newcommand*\diff{\mathop{}\!\mathrm{d}}

\newcommand{\pb}{p_{\rm branch}}
\newcommand{\pg}{p_{\rm grow}}
\newcommand{\ps}{p_{\rm stop}}

\begin{document}

\title{Minimal branching and fusion morphogenesis \\ approaches biological multi-objective optimality}

\author{Maxime Lucas}
\email{maxime.lucas@unamur.be}
\thanks{M.L. and C.B. contributed equally to this work.}
\affiliation{Department of Mathematics \& Namur Institute for Complex Systems (naXys), University of Namur, 5000 Namur, Belgium}
\affiliation{Earth and Life Institute, UCLouvain, Belgium}

\author{Corentin Bisot}
\email{C.Bisot@amolf.nl}
\thanks{M.L. and C.B. contributed equally to this work.}
\affiliation{EMBL, Heidelberg, Germany}
\affiliation{Amsterdam Institute for Life and Environment, Vrije Universiteit, Amsterdam, The Netherlands}

\author{Giovanni Petri}
\affiliation{NPLab, Network Science Institute, Northeastern University London, London, UK}
\affiliation{CENTAI Institute, Turin, Italy}

\author{Stéphane Declerck}
\affiliation{Earth and Life Institute, UCLouvain, Belgium}

\author{Timoteo Carletti}
\affiliation{Department of Mathematics \& Namur Institute for Complex Systems (naXys), UNamur, Belgium}

\date{\today}

\begin{abstract}

Many biological networks grow by elongation of filaments that can branch and fuse---typical examples include fungal mycelium or slime mold. These networks must simultaneously perform multiple tasks such as transport, exploration, and robustness under finite resources. Yet, how such multi-task architectures emerge from local growth processes remains poorly understood. Here, we introduce a minimal model of spatial network morphogenesis based solely on stochastic branching, fusion, and stopping, during elongation. Despite the absence of global optimization or feedback, the model generates a broad morphospace from tree-like, to loopy, as well as hybrid architectures. By quantifying multiple functional objectives, we show that (i) these synthetic structures occupy similar regions of performance space than evolved empirical fungal networks, and (ii) that their Pareto front of optimal trade-offs lies close to that of these same fungal networks. Our results show that biological architectures approaching multi-objective optimality can arise from simple local growth rules, and identify branching and fusion as fundamental ingredients shaping the architecture of living transport networks.

\end{abstract}

\maketitle










\section{Introduction}

Branching networks that grow by elongation of filaments are ubiquitous in biology---they appear across scales and kingdoms, from vascular and neural systems in animals to root systems, fungal mycelia, and slime mold~\cite{fricker2017mycelium,matos2025leaf,hannezo2017unifying,ronellenfitsch2016global,alim2018fluid,dorlodot2007root}. These networks must typically perform multiple functions simultaneously, including efficient transport, robustness to damage, and exploration of space, under limited resources. Fungal networks are a particularly compelling case \cite{heaton2012analysis,onnela2012taxonomies,lee2017mesoscale,aguilar-trigueros2022network,oyartegalvez2025travellingwave,voets2009extraradical, voets2006glomeraceae,chassereau2025full}: unlike vasculature or neural systems, which serve as subsystems embedded within a larger organism, the fungal network \emph{is} the organism itself, and its network architecture must  support all essential functions simultaneously~\cite{bebber2007biological}.

The structure of biological branching networks has long been studied theoretically through optimization and adaptation frameworks.
Classical theories such as Murray’s law \cite{murray1926physiological,murray1926physiological} and allometric scaling models \cite{west1999general} describe globally optimized, often tree-like architectures that minimize a single energetic objective, but they are static and preclude loops. Adaptive models driven by flow reinforcement typically start from a grid and focus on remodeling rather than growth~\cite{tero2010rules,gounaris2024braesssa,ronellenfitsch2019phenotypes}. In spatial network theory~\cite{barthelemy2022spatial,barthelemy2018morphogenesis}, most generative models connect pre-existing points---modeling for example cities that need to be reach by the rail service---rather than constructing networks through elongation-driven growth~\cite{barthelemy2025lines}. Finally, the few realistic models of fungal growth are too complex for general phenomenological analysis of their functional trade-offs~\cite{deligne2019analysis,vidal-diezdeulzurrun2017modelling}. 

Yet, elongation-based growth and loop formation are central biological features of many branching systems~\cite{zukowski2024breakthroughinduced}.
In fungal mycelia, branches elongate at their tips and frequently fuse through anastomosis~\cite{voets2006glomeraceae}, creating loops that provide alternative transport routes and resilience to local damage~\cite{giaccone2025coexistence,banavar2000topology}. Empirical networks often exhibit a coexistence of tree-like and loopy motifs~\cite{oyartegalvez2025travellingwave}. Understanding how such hybrid architectures arise from local processes, and what for what trade-off they are optimal, remains an open challenge.

Here, we introduce a minimal model of spatial network morphogenesis based solely on stochastic branching, stopping, or fusing upon spatial encounter, during branch growth.
Importantly, the model contains no global objective, no optimization, and no flow- or damage-dependent feedback. Despite its simplicity, we show that this minimal growth model (i) produces a morphospace of architectures from tree-like to loopy, that (ii)  naturally spans a morphospace of multitask network performance closely matching that of empirical fungal networks.
By quantifying transport efficiency, robustness, and space exploration, we find that the set of optimal solutions---the Pareto front~\cite{shoval2012evolutionary,santoro2018pareto}---produced by the model lies close to that of empirical filamentous fungi, and correlations between functional metrics observed in data emerge directly from the underlying growth rules. These results show that biological multi-objective optimality can approached with simple local stochastic morphogenesis, without the need for explicit optimization.

\section{Model}

\begin{figure}[tb]
\centering
\includegraphics[width=\linewidth]{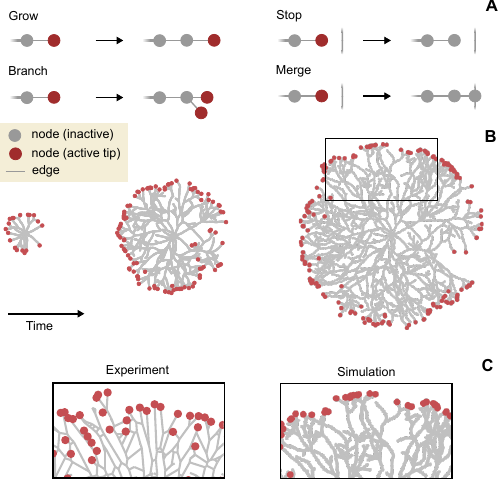}
\caption{\textbf{Minimal model of spatial network growth} \textbf{A.}~ The four local processes of the model: each active tip node (red) can grow, branch, stop, or fuse with an existing edge. \textbf{B.}~Typical network growth generated by the model. Active tips (red nodes) form a travelling front, and core edge (grey) density saturates. \textbf{C.}~Zoomed-in versions of the network from \textbf{B} (right), compared to an empirical network from a filamentous fungus~\cite{aguilar-trigueros2022network} (left).}
\label{fig:model}
\end{figure}

We now define a minimal model of spatial branching morphogenesis based on four local stochastic processes: growth, branching, stopping, andfusing (\cref{fig:model}A). The model describes the growth of a network embedded in two-dimensional continuous space and evolving in discrete time. The network is initialized as a star: a root node connected to $n$ edges terminating in $n$ active tip nodes, from which morphogenesis proceeds.
At each time step, and independently for each active tip, the network attempts an elongation event, which can be either growth or branching with respective probabilities $\pg$ and $\pb$ ($\pg+\pb=1$). Growth extends the tip by adding a single new edge of length $\delta l$ and a new active tip at its end. Branching creates two new edges of length $\delta l/2$, one continuing in the forward direction and one deviating by a fixed branching angle $\alpha$, thereby increasing the number of active tips by one.
Before any elongation occurs, the tip probes its local environment for potential intersections. If an existing edge lies along the intended elongation path at a distance $d \le k\,\delta l$, with sensing factor $k \ge 1$, the tip becomes inactive with probability $\ps$, representing growth arrest due to spatial obstruction. With complementary probability $1-\ps$, elongation proceeds as if no obstacle were present.
If an actual geometric intersection occurs during the elongation or branching process, the growing segment deterministically fuses with the encountered edge, corresponding to anastomosis. This fusion is implemented by inserting a new node at the intersection point, fusing the two edges. 
%
The growth process terminates once the total network length reaches a given value $L$. 
The model is fully determined by five parameters: two independent probabilities $\pb$ and $\ps$, a sensing factor $k$, a branching angle $\alpha$ and a unit length $\delta l$. For simplicity, we set $\alpha=\pi/6$,  $k=10$, $\delta l=1$, and vary the two probabilities. 

The networks resulting from this dynamics are spatial graphs whose edges represent filaments and whose nodes are embedded in continuous space. Nodes correspond to growth tips, filament continuations, or junctions, and are associated with degree 1, 2, and 3, respectively. Because elongation occurs in the plane and intersections lead to fusion, the resulting networks are planar by construction. Importantly, the network emerges solely from local stochastic growth, branching, stopping, and fusion rules, without any global optimization or feedback. Also note that, by design, the network is purely topological and geometric: edges are unweighted and undirected.
In \cref{fig:model}B, we illustrate a typical realization of the model. The network grows outward from the origin, forming a travelling front of active tip nodes (red), while the interior saturates to an approximately uniform density. Despite the simplicity of the model, this dynamics closely resembles empirical observations in arbuscular mycorrhizal fungi~\cite{oyartegalvez2025travellingwave}.

In this model,  loops and hence alternative pathways are formed through fusion, a feature known to be biologically relevant in fungal networks and leaf venation~\cite{matos2025leaf}. Models lacking fusion, such as Ref.~\cite{hannezo2017unifying}, correspond to the limiting case $\ps=1$ of our framework and produce purely tree-like architectures.

\section{Branching network morphospace}

We explore the morphospace of our model by systematically varying the two independent probability parameters $\pb$ and $\ps$ while fixing all other parameter values. For each parameter combination, we simulate three network realizations, each grown to the same total length $L$, which also represents the total cost of the networks.

\begin{figure}[b]
\centering
\includegraphics[width=\linewidth]{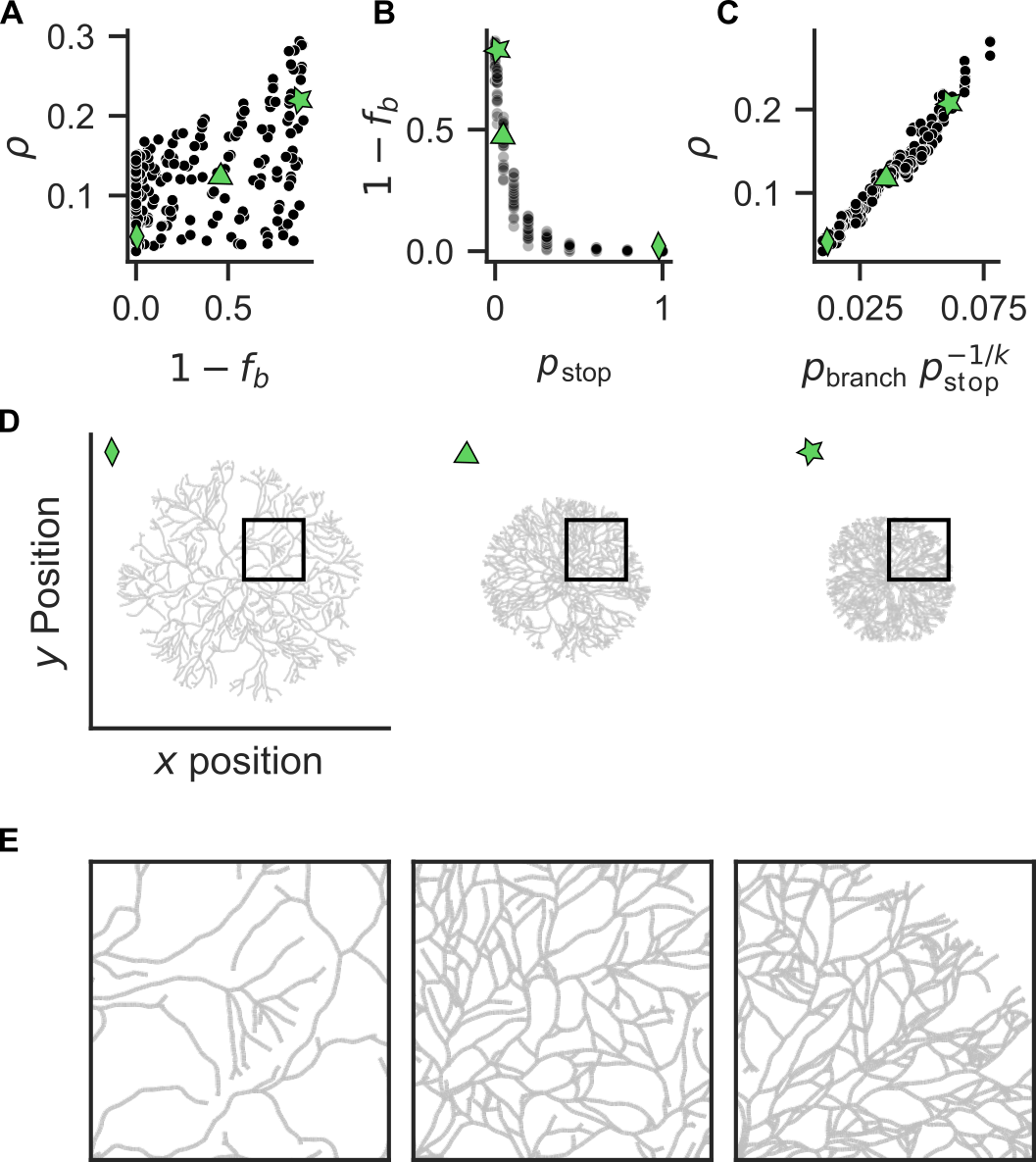}
\caption{\textbf{Morphospace of branching networks.} To explore the morphospace associated with the model, we simulate networks for a range of parameter values ($\pb$, $\ps$), with 3 realizations each. \textbf{A.}~Morphospace defined by network density $\rho$ and loopiness $1-f_b$. Each point represents one simulated network with total length $10^4$. \textbf{B, C.~} The position in this morphospace is controlled by the two parameters and $k$, the sensing distance factor. \textbf{D.~} Examples of three network architectures, from sparse and tree-like (left) to dense and loopy (right), and an intermediate hybrid architecture (middle). \textbf{E.~} Zoomed-in versions of the above networks.}
\label{fig:morphospace}
\end{figure}

\Cref{fig:morphospace}A shows the resulting morphospace in terms of network density $\rho$ (total length per unit area) and loopiness $1-f_b$ (fraction of edges in loops, similar to meshedness). Each point represents one simulated network. The morphospace spans a continuous range of architectures (\cref{fig:morphospace}D, E) from sparse tree-like structures (low $\rho$, low $1-f_b$) to dense loopy networks (high $\rho$, high $1-f_b$). 
Notably, the intermediate regime produces hybrid architectures where tree-like and loop-rich regions coexist within a single network---a structural motif commonly observed in biological systems.
This diversity  emerges from just two control parameters. 
Loopiness is determined by the stopping probability $p_{\rm stop}$: higher $p_{\rm stop}$ prevents fusion events which results in fewer loops (\cref{fig:morphospace}B). In turn, the density is proportional to $p_{\rm branch}$ but inversely proportional to $p_{\rm stop}^{1/k}$, where $k$ is the sensing distance (\cref{fig:morphospace}C).

\section{Multi-objective performance and Pareto analysis}

\begin{figure*}[htbp]
\centering
\includegraphics[width=0.99\linewidth]{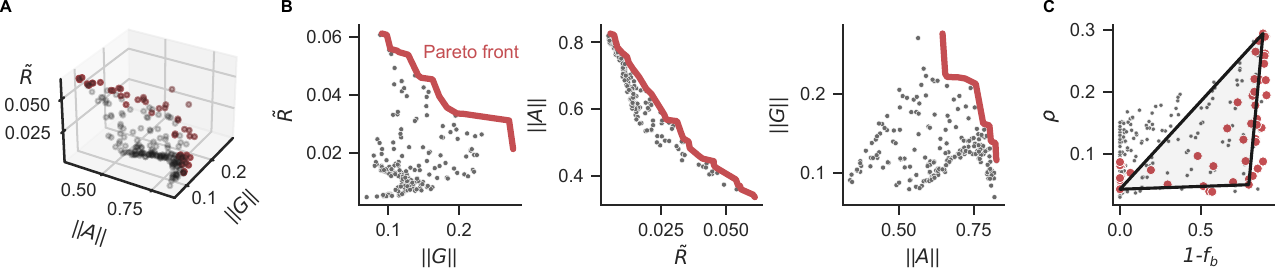}
\caption{\textbf{Multi-objective optimality with Pareto analysis.} \textbf{A.} Performance space defined by three objectives formalized by network metrics: robustness to damage $\tilde R$, space exploration $\| A \|$, and conductance $\| G \|$ for transport efficiency. Each point (grey) represents one simulated network with total length $10^4$ (same data as in \cref{fig:morphospace}A). The points fall on a curved surface. 
Networks that are non-dominated define the set of optimal solutions, called Pareto front (red).
\textbf{B.}~Two-dimensional projections of performance space.  \textbf{C.~} Visualization of the Pareto front solutions (red dots) in morphospace. The triangle (black) is the best fitting polygon following the method in~\cite{shoval2012evolutionary}.} 
\label{fig:pareto}
\end{figure*}

To assess the functional implications of this morphological diversity, we evaluate each network according to three physically interpretable performance objectives: robustness to damage, space exploration, and long-range transport. 

For biological networks, spatial exploration determines the extent of environmental interaction. We define the total area of influence $A$ by the area of the buffered network, where each edge is buffered by a width $d$. This models nutrient uptake: for $d = \sqrt{Dt}$ where $D$ is solute diffusivity and $t$ is a depletion timescale, $A$ is the area from which nutrients can be absorbed. To ensure scale-invariance, we normalize it by the maximum possible influence area $A_{\max} = 2Ld$, defining the \textit{exploration metric} in $[0, 1]$
\begin{equation}
    \|A\| = \frac{A}{2Ld}.
\end{equation}
Note that the choice of $d$ is important: values too small will yield $\|A\| \approx 1$ because no buffered edges overlap, whereas values too large yield $\|A\| \approx 0$ because of maximum overlap. Intermediate values must be chosen. 

Biological networks must also withstand localized structural damage from grazing or mechanical breakage, for example. We model damage by removing edges randomly until a fraction $f$ of total length $L$ is removed. For the damaged network, we measure $L_{\rm LCC}(f)$, the total length of the largest connected component. The \textit{robustness metric} is then defined in $[0,1]$ as the area under the curve
\begin{equation}
    R = \int_0^1 \left\langle \frac{L_{\rm LCC}(f)}{L} \right\rangle \diff \! f,
\end{equation}
where $\langle \cdot \rangle$ denotes averaging over multiple damage realizations. This quantifies the network's ability to maintain connectivity under progressive damage: larger values indicate a more robust network. 

We treat the network as resistors with conductances inversely proportional to edge lengths and compute the total conductance $G$ between opposite network boundaries. To compare networks of different spatial extents, we note that rescaling a network's linear dimensions by factor $\beta$ (while preserving total material) scales conductance as $\beta^{-3}$. We therefore normalize by spatial extent (measured by influence area $A$) and introduce
\begin{equation}
    \tilde G = G \cdot \| A \|^3,
\end{equation}
further normalized to $[0,1]$ using the theoretical minimum and maximum conductances for a network of length $L$ (see Supplementary Material):
\begin{equation}
||G|| = \frac{\tilde{G}-\tilde{G}_{\rm min}}{\tilde{G}_{\rm max}-\tilde{G}_{\rm min}} .
\end{equation}

\Cref{fig:pareto}A shows the resulting three-dimensional performance space. Each point represents one simulated network, positioned according to its performance on all three objectives. The networks span a curved surface in this space, revealing inherent trade-offs: networks that are optimal for one objective necessarily sacrifice performance in others. For instance, tree-like networks excel at space exploration but are fragile and have poor transport efficiency, while loopy networks are robust and efficient transporters but explore space less effectively.
The two-dimensional projections (\cref{fig:pareto}B) reveal the structure of these trade-offs more clearly. In each projection, we identify the Pareto front (red line)---the set of non-dominated solutions where no network performs better on all objectives simultaneously. Networks on this front represent architectures with optimal trade-offs between competing objectives. 

We then map the performance space and the Pareto front back onto the morphospace (\cref{fig:pareto}C). 
Importantly, the Pareto front is not restricted to extreme morphologies: intermediate structures with hybrid tree-loop architecture occupy the central regions of the front, indicating that they achieve balanced multi-objective performance.
This also shows that the two parameters $\pb$ and $\ps$ can tune network architecture along the multi-objective trade-off surface.

\section{Synthetic networks approach empirical multi-objective optimality}

\begin{figure}[b]
\centering
\includegraphics[width=\linewidth]{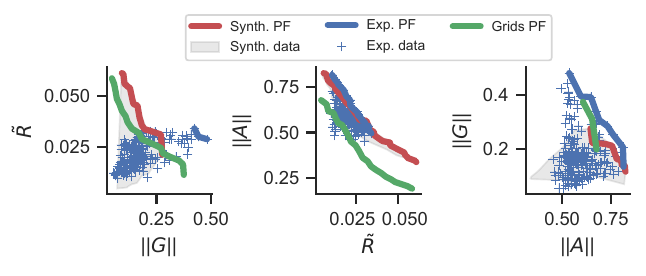}
\caption{\textbf{Model-grown networks approach empirical multi-objective optimality.} In the same 2D performance spaces as in \cref{fig:pareto}, we show the region occupied by simulated networks (grey shade), the associated Pareto front (PF), the 253 empirical networks (blue crosses) and their PF (blue line), and the grids' PF. The PF front of the simulated networks lies close to the other two. }
\label{fig:data}
\end{figure}

Having characterized the performance landscape accessible to our growth model, we now ask: how do these synthetic networks compare to biological networks shaped by evolution?
To test this, we compare our model-generated networks (grey shaded region) to empirical data (blue crosses) from 253 fungal networks from Ref.~\cite{aguilar-trigueros2022network} (\cref{fig:data}) in the same 2D-projected performance spaces as in \cref{fig:pareto}. 

The key observation is that synthetic networks generated by our simple stochastic growth rules closely approach the performance levels of biological networks. Indeed, the synthetic Pareto front (red) lies near the empirical Pareto front (blue line), and many individual fungal networks fall within or near the region occupied by synthetic networks. This suggests that the growth dynamics captured by our minimal model---stochastic branching and fusion---are sufficient to generate network architectures with near-biological performance, without requiring adaptive reinforcement or global optimization processes. We push the comparison further with extremes structures: grid networks. Grids, by construction, are expected to be very robust but less efficient at exploration. Interestingly, their Pareto front (green) lies near those of both our model's and empirical networks. This suggests that the flexibility of growth-based morphogenesis enables exploration of a richer performance landscape than static regular structures.

Notably, the empirical fungal networks span a similar range of trade-offs as our synthetic networks, from tree-like explorers to loopy transporters. This morphological diversity in nature mirrors the diversity generated by varying our model's stopping probability $p_{\rm stop}$, supporting the hypothesis that evolution may tune network architecture primarily through relatively simple parameters controlling local growth rules rather than implementing complex global optimization strategies.

\section{Discussion}

We have shown that minimal local growth rules-stochastic branching and fusion—are sufficient to generate spatial networks with multitask performance approaching that of evolved empirical fungal networks. Without global or local optimization, simple morphogenesis produces a wide morphological diversity observed in biological branching networks, from tree-like explorers to loopy transporters, by tuning two physical parameters. 

Interestingly, intermediate parameter values produce hybrid structures where tree-like and loop-rich regions coexist. This coexistence is often observed in biological systems, but has usually been explained as a response to damage or fluctuating flows \cite{katifori2010damage}. 

Our framework is generally applicable to branching systems that grow via the elongation of their branches---root networks, vasculature, urban infrastructure---where the interplay between local growth rules and global performance constraints the emergent architecture. More broadly, this work demonstrates that complex multitask optimization can emerge from simple stochastic processes, a principle relevant beyond biology to any system where structure grows through local interactions.

Several interesting questions remain open. First, while our minimal model captures core morphogenetic processes, real biological systems exhibit additional complexity: for example, neurons distinguish deterministic primary branching from stochastic secondary elaboration~\cite{ronellenfitsch2016global}, vascular networks undergo pruning and reinforcement driven by flow feedback~\cite{hu2013adaptation,tero2010rules}. How do these additional mechanisms alter the accessible morphospace and shift the Pareto frontier? A systematic comparison of growth models with varying levels of complexity could reveal which ingredients are essential for biological performance and which provide marginal refinements. Second, our planar networks neglect the third dimension available to real mycelia, root systems, and vasculature. Three-dimensional growth may relax topological constraints---allowing branches to cross withoutfusioning---and thereby access novel regions of performance space. Finally, we hope this work will stimulate others studies of elongation-based growth models of spatial networks which are still lacking~\cite{barthelemy2025lines}.


\bibliography{bib}

\begin{acknowledgments}
The authors would like to thank Carlos Aguilar-Trigueros for making his data available. 
M.L. is a Postdoctoral Researcher of the Fonds de la Recherche Scientifique–FNRS.
\end{acknowledgments}

\noindent \textbf{Data availability:} 
The data used in this study comes from~\cite{aguilar-trigueros2022network} and is available at https://zenodo.org/records/14931953.

\noindent \textbf{Code availability:} 
Code for reproducing our results will be made available online upon publication.


\noindent \textbf{Competing interests:} 
The authors declare no competing interests.

\clearpage

\setcounter{figure}{0}
\setcounter{table}{0}
\setcounter{equation}{0}
\setcounter{page}{1}
\setcounter{section}{0}

\makeatletter
\renewcommand{\thefigure}{S\arabic{figure}}
\renewcommand{\theequation}{S\arabic{equation}}
\renewcommand{\thetable}{S\arabic{table}}

\setcounter{secnumdepth}{2} 


    


\end{document}